\documentclass[prd,twocolumn,tightenlines,showpacs,nofootinbib,superscriptaddress]{revtex4-1} 

\usepackage{amsfonts}
\usepackage{dcolumn}
\usepackage{bm}
\usepackage{amsmath}
\usepackage{nicefrac}
\usepackage{amsthm}
\usepackage{amssymb}
\usepackage{graphicx}
\usepackage{color}

\newcommand{\appropto}{\mathrel{\vcenter{
  \offinterlineskip\halign{\hfil$##$\cr
    \propto\cr\noalign{\kern2pt}\sim\cr\noalign{\kern-2pt}}}}}

\renewcommand{\v}[1]{\boldsymbol{#1}}		

\begin{document}

\title{Comment on ``Axion induced oscillating electric dipole moments''}

\date{\today}
\author{V.~V.~Flambaum} 
\affiliation{School of Physics, University of New South Wales, Sydney 2052, Australia}
\affiliation{Mainz Institute for Theoretical Physics, Johannes Gutenberg University Mainz, D 55122 Mainz, Germany}
\author{B.~M.~Roberts} 
\affiliation{School of Physics, University of New South Wales, Sydney 2052, Australia}
\author{Y.~V.~Stadnik} 
\affiliation{School of Physics, University of New South Wales, Sydney 2052, Australia}

\begin{abstract}
In the recent work [Phys.~Rev.~D \textbf{91}, 111702(R) (2015)], C.~Hill concludes that the axion electromagnetic anomaly induces an oscillating electron electric dipole moment of frequency $m_a$ and strength $\sim 10^{-32}~e$ cm, in the limit $v/c \to 0$ for the axion field. Here, we demonstrate that a proper treatment of this problem in the lowest order yields \emph{no} electric dipole moment of the electron in the same limit. Instead, oscillating electric dipole moments of atoms and molecules are produced by different mechanisms. 
\end{abstract}


\maketitle 

In Ref.~\cite{Hill2015}, it is concluded that the axion electromagnetic anomaly induces an oscillating electron electric dipole moment (EDM) of frequency $m_a$ and strength $\sim 10^{-32}~e$ cm, in the limit $v/c \to 0$ for the axion field, which is several orders of magnitude stronger than the axion-induced oscillating EDMs of nucleons induced by the Quantum Chromodynamics (QCD) anomaly \cite{Graham2011}, and the axion-induced oscillating EDMs of atoms due to the QCD anomaly, as well as the derivative interaction of the axion field with atomic electrons \cite{Stadnik2014axions,Roberts2014prl,Roberts2014long}. We disagree with Ref.~\cite{Hill2015} for the following reasons:

(1) The EDM of an elementary particle is defined by the linear energy shift that it produces through its interaction with a \emph{static} applied electric field:~$\delta \varepsilon = -\v{d} \cdot \v{E}$. 
As we show below, the interaction of an electron with a static electric field, in the presence of the axion electromagnetic anomaly, in the lowest order does not produce an energy shift in the limit $v/c \to 0$. 
This implies that \emph{no} electron EDM is generated by this mechanism \cite{Footnote1}.

(2) In the non-relativistic limit, the electron EDM is fully screened in a neutral composite system \cite{Schiff1963}. Indeed, it is the atomic or molecular EDM that is actually measured in an experiment. This induced EDM appears due to the relativistic corrections for the interaction of the electron EDM with the  atomic/molecular electric field \cite{Sandars1965,Flambaum1976}. Such an interaction with the static electric field vanishes within the approach of Ref.~\cite{Hill2015}. Therefore, a fully relativistic treatment for both the electron EDM and atomic/molecular EDMs induced by the axion field is necessary.

\begin{figure}[h!]
\begin{center}
\includegraphics[width=4cm]{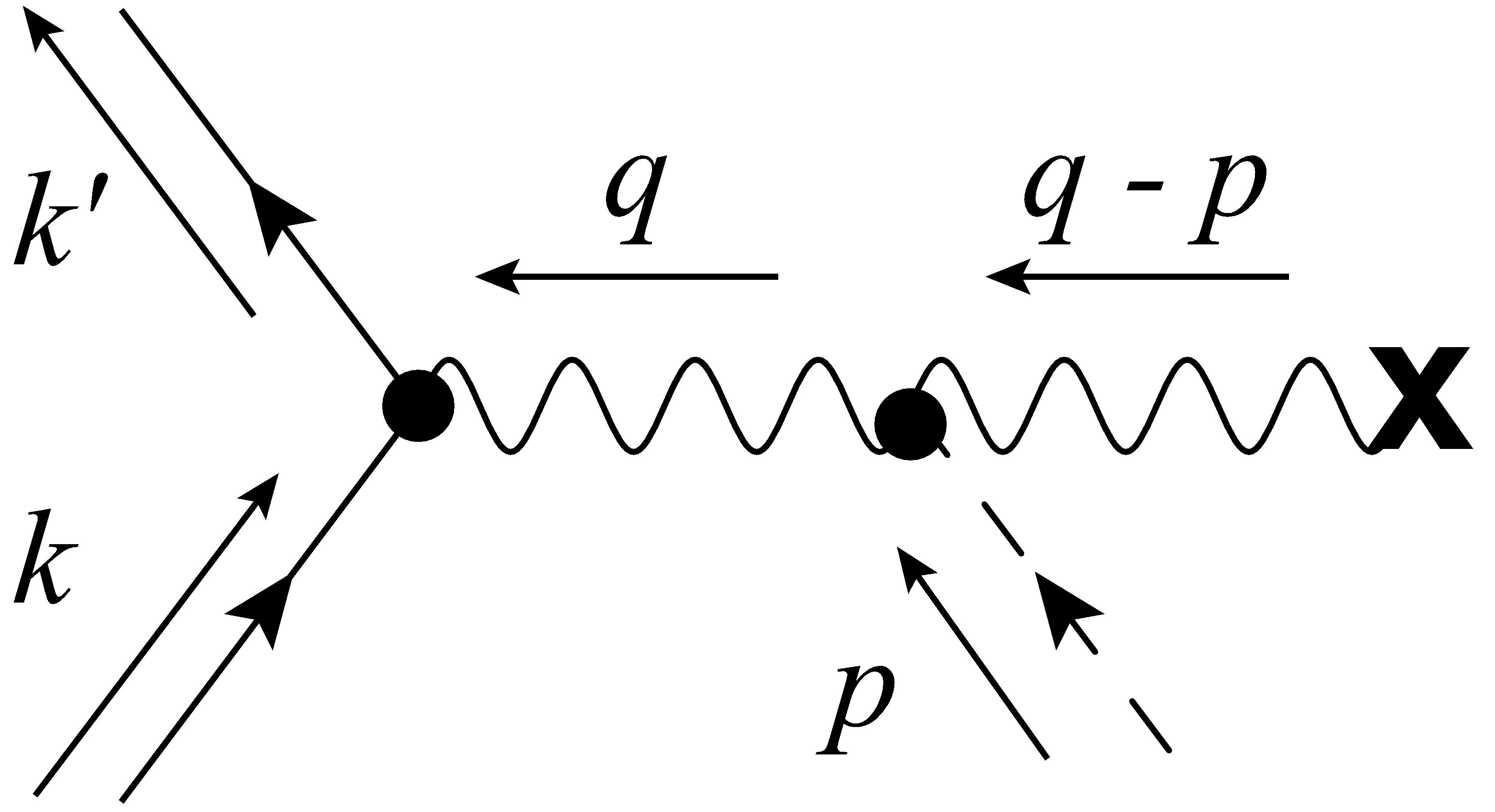}
\caption{Feynman diagram for the interaction of an electron with a static electric field (bold cross), in the presence of the axion electromagnetic anomaly.} 
\label{fig:electron_axion_anomaly_EDM}
\end{center}
\end{figure}

In the limit $v/c \to 0$, the interaction of an electron with a static applied electric field, in the presence of the axion electromagnetic anomaly due to the axion field $a(t) = a_0 \cos(m_a t)$ (represented by the Feynman diagram in Fig.~\ref{fig:electron_axion_anomaly_EDM}), has the amplitude:
\begin{align}
\label{d_edm}
\mathcal{M} = - \frac{e  g_{a\gamma\gamma} a_0 \sin(m_a t)}{f_a} \frac{\varepsilon_{\nu \rho \sigma \mu} p^\nu \varepsilon^\rho q^\sigma \bar{u}(k') \Gamma^\mu u(k)}{ q^2} ,
\end{align}
with the axion 4-momentum $p^\nu = (m_a, \v{0})$ and the dressed electromagnetic vertex $\Gamma^\mu = F_1(q^2) \gamma^\mu + i F_2(q^2) \sigma^{\mu \nu} q_\nu / 2m$, where $F_1$ and $F_2$ are the Dirac and Pauli form factors, respectively, of the electron. 
Note that the energy shift and applied electric field are gauge-invariant quantities, meaning that the EDM is also gauge-invariant. 
Since the axion-induced EDM is independent of the choice of gauge, we choose the simplest gauge: $V = -\v{E} \cdot \v{r}$, $\v{A} = \v{0}$, in which the polarisation vector is purely timelike: $\varepsilon^\rho = (\varepsilon^0, \mathbf{0})$. In this gauge, it is evident that two of the indices in the antisymmetric Levi-Cevita tensor $\varepsilon_{\nu \rho \sigma \mu}$ in Eq.~(\ref{d_edm}) are forced to be equal ($\nu = \rho = 0$), meaning that the amplitude associated with the process in Fig.~\ref{fig:electron_axion_anomaly_EDM}, as well as the resulting EDM of the electron, both vanish in the limit $v/c \to 0$:
\begin{align}
\label{d_edm_final}
\mathcal{M} = 0 ~ => ~ d_e = 0 .
\end{align}
In the case of a slowly changing applied electric field, the 4-potential may be presented as $A^\mu = (V,\v{A})$, where $V(\v{r},t) = -\v{E} (t) \cdot \v{r}$, $\v{A}(\v{r},t) = - (d\v{E}(t) / dt) r^2$/4.  For an oscillating applied electric field, corrections are proportional to the electric field oscillation frequency $\omega$ and are suppressed by the small parameter $\omega/m_e$.

The fact that the effect is zero for the static electric field also follows from Eq.~(5) in Ref.~\cite{Hill2015}.
Indeed, the virtual photons for a static electric field are longitudinal, with the polarisation vector $\varepsilon$ directed along the momentum $\v{k}_f$ (since the electric force and the momentum transfer $\v{k}_f$ are directed along the electric field, contrary to the free photon case where the polarisation vector $\varepsilon$ is perpendicular to $\v{k}_f$). In this case, the antisymmetric tensor $\varepsilon_{\nu \rho \sigma \mu}\varepsilon^{\nu} k_f^{\rho}$ vanishes. The calculation with a transverse photon presented in Ref.~\cite{Hill2015} actually corresponds to the conversion of an axion to a photon in the magnetic field of an electron (which has no relation to the electron EDM).


However, axions may induce oscillating EDMs in paramagnetic atoms and molecules through three possible mechanisms (mechanisms (I) and (II) have been investigated in Refs.~\cite{Stadnik2014axions,Roberts2014prl,Roberts2014long}):

(I) Axion-induced oscillating nuclear Schiff and magnetic quadrupole moments, which are induced primarily through the $P$,$T$-violating $NN$ interaction mediated by $\pi$ exchange, and also through the intrinsic nucleon EDMs, both due to the QCD anomaly. 

(II) The derivative interaction of the axion field with atomic/molecular electrons.

\begin{figure}[h!]
\begin{center}
\includegraphics[width=5cm]{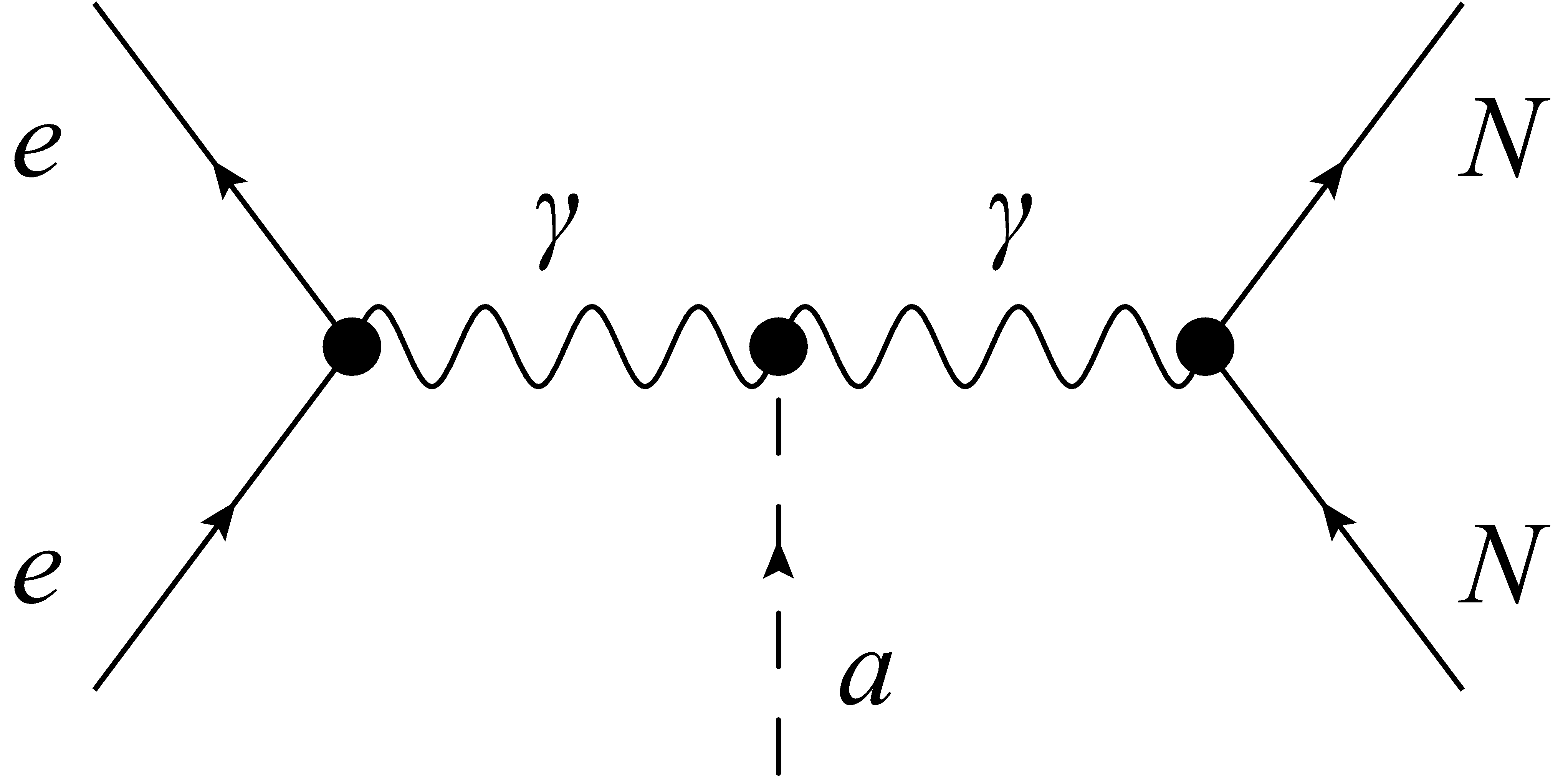}
\caption{Feynman diagram for the  interaction of atomic/molecular electrons and nucleons, perturbed by the axion electromagnetic anomaly.} 
\label{fig:Axion-electron-nucleon_anomalous_2}
\end{center}
\end{figure}

(III) Perturbation of the electron-nucleon Coulomb interaction by the axion electromagnetic anomaly (Fig.~\ref{fig:Axion-electron-nucleon_anomalous_2}).

\textbf{Acknowledgments} --- This work was supported by the Australian Research Council. B.~M.~R.~and V.~V.~F.~are grateful to the Mainz Institute for Theoretical Physics (MITP) for its hospitality and support.



\end{document}